\documentclass[5p,twocolumn,sort&compress]{elsarticle}
\usepackage{amssymb}
\usepackage{amsmath}
\usepackage{hyperref}
\usepackage{physics}
\usepackage{hubbleflow}
\usepackage{xcolor}
\usepackage{subcaption}

\newcommand{\calZ}{\mathcal{Z}}

\DeclareMathOperator{\Ei}{Ei}
\DeclareMathOperator{\Eone}{E_1}

\journal{Physics of the Dark Universe}

\begin{document}

\begin{frontmatter}

  \title{Comment on: ``Third-order corrections to the slow-roll
    expansion: Calculation and constraints with Planck, ACT, SPT, and
    BICEP/Keck [2025 PDU 47 101813]''}
  \author[1]{Pierre Auclair}
  \ead{auclair@iap.fr}
  \author[2]{Christophe Ringeval}
  \ead{christophe.ringeval@uclouvain.be}
  
  \affiliation[1]{organization={Institut d'Astrophysique de Paris},
    addressline={98bis Boulevard Arago}, 
    city={Paris},
    postcode={75014},
    country={France}}

  \affiliation[2]{organization={Cosmology, Universe and Relativity at Louvain (CURL),
      Institute of Mathematics and Physics, University of Louvain},
    addressline={2 Chemin du Cyclotron}, 
    city={Louvain-la-Neuve},
    postcode={1348},
    country={Belgium}}

  \begin{abstract}
    We point out that several terms in the third-order corrections to
    the slow-roll power spectra presented by
    \citet{Ballardini:2024irx} are incorrect. The authors of that work
    present their result as differing from the ones originally presented
    by \citet{Auclair:2022yxs} due to some different approximation
    schemes. However, in our original work, all terms at all orders
    have been derived exactly and any difference between two
    expansions performed at the same pivot wavenumber signals a
    problem. As we show in this comment, \citet{Ballardini:2024irx}
    have misevaluated some definite three-dimensional integrals by
    integrating a truncated Taylor expansion instead of Taylor expanding an
    integral. Our claim is backed-up with a Monte-Carlo numerical
    integration of the incriminated three-dimensional integrals,
    which, unsurprisingly, matches the analytical value derived in
    \citet{Auclair:2022yxs}.
  \end{abstract}

  
\end{frontmatter}

\section{Introduction}

Cosmic inflation, an early era of accelerated expansion of the
spacetime, is one of the most compelling paradigms to solve various
problems of the Friedmann-Lema\^{\i}tre hot Big-Bang model while
providing a physical mechanism at the origin of the cosmic microwave
background anisotropies and of the large scale structures of the
universe~\cite{Starobinsky:1979ty, Starobinsky:1980te, Guth:1980zm,
  Linde:1981mu, Albrecht:1982wi, Linde:1983gd, Mukhanov:1981xt,
  Mukhanov:1982nu, Starobinsky:1982ee, Guth:1982ec, Hawking:1982cz,
  Bardeen1983}. In its simplest incarnation, cosmic inflation can be
realized by a single scalar field slowly rolling down its potential,
the quantum fluctuations of which acting as the seeds of all the
structures observed today. Single-field inflation is a landscape
populated by hundreds of different models that have been put to the
test with ever-increasingly accurate cosmological
data~\cite{Martin:2013tda, EI:Opiparous, Martin:2024qnn,
  Martin:2024nlo, Ringeval:2025vus}. In order to predict the actual
shape of the primordial scalar and tensor power spectra associated
with any inflationary models, one can rely on exact numerical
integrations~\cite{Salopek:1998qh,Adams:2001vc,Ringeval:2007am,Mortonson:2010er,Seery:2016lko,
  Werth:2024aui}. However, for the single-field slow-roll models,
there exists a unified framework, based on a perturbative expansion,
that allows us to derive the functional shape of the power
spectra~\cite{Stewart:1993bc, Liddle:1994dx, Nakamura:1996da,
  Gong:2001he, Hoffman:2000ue, Schwarz:2001vv, Leach:2002ar,
  Schwarz:2004tz}. These expansions have been shown to be very
precise~\cite{Makarov:2005uh,Ringeval:2013lea,Martin:2016iqo,Auclair:2024udj}
and their accuracy can always be increased by pushing them to higher
orders. They are constructed over a hierarchy of the so-called
``Hubble-flow'' functions defined by
\begin{equation}
\eps{i+1}(N) \equiv \dv{\ln |\eps{i}|}{N}, \qquad
\eps{1}(N) \equiv -\dv{\ln H}{N}\,.
\label{eq:hubbleflow}
\end{equation}
Here $N=\ln a$ stands for the number of e-folds, $a(\eta)$ being the
Friedmann-Lema\^{\i}tre scale factor while $H(N)$ is the Hubble
parameter during inflation, which is almost constant for quasi-de
Sitter spacetime thereby ensuring the smallness of the $\eps{i}$
functions.

Motivated by the soon-to-be released Euclid satellite
data~\cite{Lacasa:2019flz, Euclid:2021qvm, Euclid:2024yrr}, we have
pushed the Hubble-flow expansion of the scalar and tensor power
spectra to third order (N3LO) in \citet{Auclair:2022yxs}. Our result
is based on the so-called Green's functions method introduced by
\citet{Gong:2001he} and it applies to all single-field slow-roll
models having minimal and non-minimal kinetic terms. Later on, our
results have been recovered and extended to other effective field
theories, including a broader class of modified gravity models, by
\citet{Bianchi:2024qyp}.

In simple terms, one perturbatively solves the evolution of
cosmological perturbations during inflation, starting from initial
conditions set by quantum fluctuations in the Bunch-Davis vacuum. By
doing so, one can finally derive the scalar power spectrum,
$\calPz(k)$, of the comoving curvature perturbation and the tensor
power spectrum, $\calPh(k)$, for the primordial gravitational
waves. Their final expressions, at third order, take the form
\begin{equation}
\begin{aligned}
  \calPz(k) & = \dfrac{\Hdiam^2}{8 \pi^2 \Mpl^2 \epsdiam{1}
  \csdiam}\left[\bsdiam{0} + \bsdiam{1} \ln
  \left(\dfrac{k}{\kdiam}\right) + \bsdiam{2}
  \ln^2\left(\dfrac{k}{\kdiam}\right) \right. \\ & + \left. \bsdiam{3}
  \ln^3\left(\dfrac{k}{\kdiam}\right) \right],
\end{aligned}
\label{eq:calPz}
\end{equation}
and
\begin{equation}
  \begin{aligned}
  \calPh(k) & = \dfrac{2 \Hdiam^2}{\pi^2 \Mpl^2}\left[\btdiam{0} + \btdiam{1} \ln
  \left(\dfrac{k}{\kdiam}\right) + \btdiam{2}
  \ln^2\left(\dfrac{k}{\kdiam}\right) \right. \\ & + \left. \btdiam{3}
  \ln^3\left(\dfrac{k}{\kdiam}\right) \right],\\  
\end{aligned}
\label{eq:calPh}
\end{equation}
where the $\bsdiam{i}$ and $\btdiam{i}$ are determined functionals of
the Hubble-flow functions $\epsdiam{i}\equiv\eps{i}(\etadiam)$
evaluated at a precise time $\etadiam$. This time corresponds to the
instant at which the reference pivot wavenumber $\kdiam$ (usually set
at $0.05\,\Mpc^{-1}$), appearing in Eqs.~\eqref{eq:calPz} and
\eqref{eq:calPh}, crosses the sound radius during inflation, i.e., it
is solution of $\kdiam \etadiam \cs(\etadiam) = -1$, $\cs(\eta)$ being
the sound speed of the cosmological perturbations during (K-)
inflation\footnote{The functionals $\bsdiam{i}$ and $\btdiam{i}$ are
also dependent on a hierarchy of sound-flow functions $\del{i}(\etadiam)$  built upon
$\cs(\eta)$, but this is not important for the present
discussion.}. Most of the action in these expansions consists in
determining all the $\bsdiam{i}$ and $\btdiam{i}$ functionals and
their (long) expression can be found in Ref.~\cite{Auclair:2022yxs} as
Eqs.~(54), and (74) to (77).

\section{Different functionals}

\begin{figure*}[h!]
  \begin{subfigure}{0.49\textwidth}
    \includegraphics[width=\textwidth]{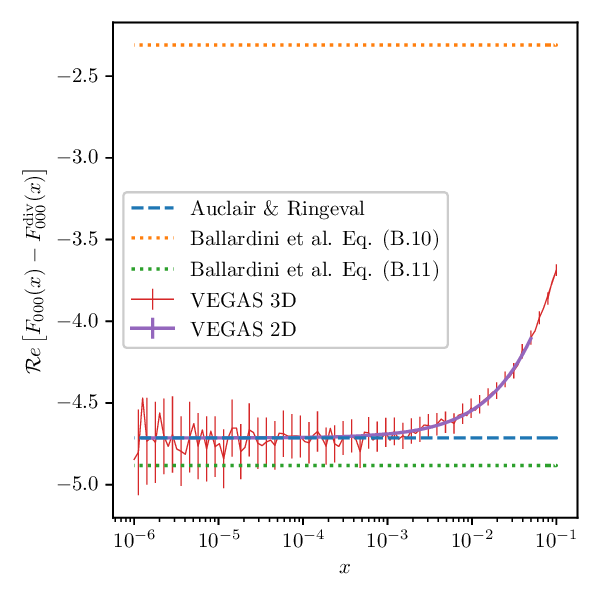}
    \caption{Real part.}
  \end{subfigure}
  \begin{subfigure}{0.49\textwidth}
    \includegraphics[width=\textwidth]{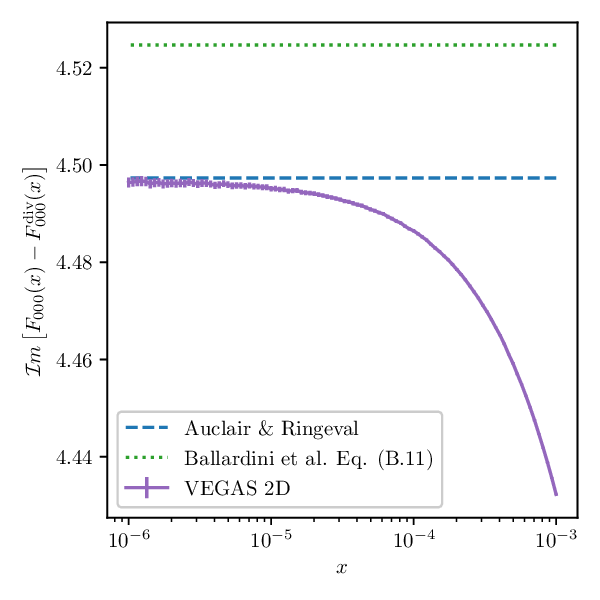}
    \caption{Imaginary part.}
  \end{subfigure}
  \caption{ Finite part of $\F{000}(x)$ computed using the VEGAS
    algorithm over a two-dimensional and a three-dimensional domain.
    All the points in the curves were computed using ten iterations of $10^8$
    samples each. The error bars show five times the estimated standard
    deviation.}
  \label{fig:divergent}
\end{figure*}

Two years after Ref.~\cite{Auclair:2022yxs},
\citet{Ballardini:2024irx} have presented ``an independent approach to
the solution of the integrals compared to the one previously presented
in the literature'', in which they have attempted to reproduce our
results in the particular case where $\cs=1$, i.e., for minimal
kinetic terms. Although most of the calculations appear to be very
similar to our previously published results, including the usage of
Green's functions, some of the $\bsdiam{i}$ and $\btdiam{i}$
functionals end up differing when compared to ours.

Independently of the complexity of the underlying calculations,
Eqs.~\eqref{eq:calPz} and \eqref{eq:calPh} are just Taylor expansions
in $\ln(k/\kdiam)$. As such, the ``coefficients'' $\bsdiam{i}$ and
$\btdiam{i}$ are universal and should numerically match. As a matter
of fact, and at least up to second order, one can check that other
approximate solving methods, such as the uniform approximation, indeed
lead to coefficients having numerical values very close to the ones
obtained with Green's functions~\cite{Martin:2002vn, Lorenz:2008et}.

\citet{Ballardini:2024irx} find a different
expression than ours for $\bsdiam{0}$ and $\btdiam{0}$ in various
coefficients multiplying terms of order three
$\order{\eps{i}^3}$. The differences find their root in the evaluation
of triple integrals of the form
\begin{equation}
  \F{000}(x) = \int_x^\infty \dfrac{e^{+2iy}}{y}\F{00}(y)\dd{y},
  \label{eq:F000}
\end{equation}
where
\begin{equation}
    \F{00}(x) = \int_x^\infty \dfrac{e^{-2iy}}{y} \F{0}(y) \dd{y},
  \label{eq:F00}
\end{equation}
and
\begin{equation}
    \F{0}(x) \equiv \int_x^\infty \dfrac{e^{+2 i y}}{y} \dd{y}.
    \label{eq:F0}
\end{equation}
In Ref.~\cite{Auclair:2022yxs}, we have found a generating functional
for all integrals of the form $\F{0^n}$ and this has allowed us to
evaluate $\F{000}(x)$ exactly (see appendix A of that
reference). Indeed, defining
\begin{equation}
\Ig(\nu,x) \equiv \sum_{k=0}^{+\infty} \I{k}(x) \nu^k,
\label{eq:Igdef}
\end{equation}
where
\begin{equation}
\I{2n}(x) = \conj{\F{0^{2n}}}(x), \qquad \I{2n+1}(x) = \F{0^{2n+1}}(x),
\label{eq:I2F}
\end{equation}
we have proven that
\begin{equation}
\begin{aligned}
  \Ig(\nu,x) & =  -x e^{ix} \left\{ \sin\left(\dfrac{\pi \nu}{2}
  \right) \left[\sbesselj{\nu}(x) + i \sbesselj{\nu-1}(x) \right]
  \right. \\ & \left. +
  \cos\left(\dfrac{\pi \nu}{2}\right) \left[\sbessely{\nu}(x) + i
    \sbessely{\nu-1}(x) \right] \right\}.
\end{aligned}
\label{eq:Igexact}
\end{equation}
In the previous expressions, $\sbesselj{\nu}(x)$ and
$\sbessely{\nu}(x)$ are the spherical Bessel functions of order $\nu$
and $\conj{\F{0^{2n}}}$ stands for the complex conjugate of $\F{0^{2n}}$. One
therefore sees that all the $\F{0^n}$ integrals can be exactly
obtained by successive differentiation of Eq.~\eqref{eq:Igexact} with
respect to $\nu$.

In fact, for the Hubble-flow expansion we are concerned with, one only
needs the limit at small $x$ of these integrals, and we find
\begin{equation}
  \begin{aligned}
\F{000}(x) & = - \frac{7}{3}  \zeta(3) -\frac{\pi^2}{4} B - \frac{1}{6}
B^3 - \left(\frac{\pi^2}{4} +  \frac{B^2}{2}\right) \ln(x) \\ &
- \frac{B}{2} \ln^2(x) - \frac{1}{6} \ln^3(x) + \order{x},
  \end{aligned}
\label{eq:exactF000}
\end{equation}
where the constant $B=\gammaE + \ln(2) - \dfrac{i \pi}{2}$, $\gammaE$ being the
Euler-Mascheroni constant.

\citet{Ballardini:2024irx} propose two methods in their appendix B to
find the limit at $x \to 0$ of $\F{000}(x)$:

\begin{itemize}
  \item The first method is described by the authors below their
    Eq.~(B.9):
    \begin{quote}
      ``to solve the triple integral entering eq.~(32), we
    take the limit for $u \to 0$ of the double integral in the
    integrand, which is eq.~(B.6), and then we integrate the leading
    contributions of order $\order{x^0}$, multiplied by $e^{2iu} /u$,
    between $x$ and $\infty$.''
    \end{quote}
    In other words, \citet{Ballardini:2024irx} integrated a truncated
    Taylor expansion around $0$ up to infinity.
  \item The second method is described above their Eq.~(B.11):
    \begin{quote}
      ``We have divided the integral into two parts: one from $x$ to $1$
      and one from $1$ to $+\infty$. The contribution of the integral
      from $1$ to $+\infty$ is negligible because the integrand $f
      (u)$ decreases rapidly for $u \gg 1$.''
      \end{quote}
\end{itemize}
For both methods, they analytically perform \emph{the integral of a
truncated Taylor expansion} of $\F{00}(y)$ at small $y$, either on the
full domain -- therefore outside the domain of validity of the
expansion -- or only within $[x,1]$, the contribution coming from the
other domain being neglected.  Let us note that the integral from $x$
to $1$ diverges logarithmically as $x \to 0$.  As a consequence, even
if the integral from $]1, \infty[$ might appear negligeable in
    comparison, it is still relevant to determine the finite part of
    $\F{000}(x)$. Moreover, integrals may have order unity values even
    when their integrand decreases rapidly.

Their results end up differing from Eq.~\eqref{eq:exactF000} by a
constant and instead of having the term in $7 \zeta(3)/3 \simeq
2.8048$ they obtain either $\calZ=\zeta(3)/3 \simeq 0.4007$ or a
precise complex number reported to be $\calZ \simeq 2.97353 -
0.0273557i$ (see Appendix B of Ref.~\cite{Ballardini:2024irx}).

In various places of their paper, \citet{Ballardini:2024irx} suggest
that the actual value of $\calZ$ is unknown and dependent on some
approximation scheme chosen to evaluate $\F{000}(x)$, our result being
only one approximation among others. Indeed, one may read on page 2
of \citet{Ballardini:2024irx}:
\begin{quote}
  ``While these results have been already
  presented by Auclair and Ringeval, we obtain them with a different
  approach to the integrals.''
\end{quote}
Then, on page 5, they write:
\begin{quote}
  ``$\calZ$ is a
  constant encoding the difference due to the approximation scheme used
  to calculate the triple integral appearing in Eq.~(32), see appendix
  B''.
\end{quote}
Let us stress that this is not the case, the \emph{exact value} of $\calZ$
is the one we have derived, namely $7 \zeta(3)/3$, simply because we
have calculated exactly the integral and, then, taken its limit at
small $x$. In other words, we have performed a Taylor expansion of an
exact expression for the integral whereas \citet{Ballardini:2024irx}
have performed an integral of a truncated Taylor expansion.

Eventually, the authors of Ref.~\cite{Ballardini:2024irx} write on
page 20:
\begin{quote}
  ``Different choices of $\calZ$ do not lead to numerically
  significant differences in the final PPS''.
\end{quote}
First, we stress that $\calZ$ is a known number, it is not a
``choice''. Second, the error on $\calZ$ concerns only third-order
terms, scaling as $\order{\eps{i}^3}$, and, indeed, changing these
terms by order unity cannot drastically change the final shape of the
power spectra. However, third-order corrections are mathematically
well-defined and relevant for future observations. What would be the
purpose of deriving third-order corrections in the first place if one
does not pay attention to their actual value?

\section{Numerical evaluation of the integrals}

We present in this section a numerical integration of $\F{000}(x)$,
defined by Eqs.~\eqref{eq:F000} to \eqref{eq:F0}. One has to perform a
three-dimensional integral of a rapidly oscillating function, from
close to zero to infinity, which is a non-trivial technical
problem. For this purpose, we have used the VEGAS algorithm, an
adaptive multidimensional Monte Carlo integration package now adapted
to Python~\cite{Lepage:1977sw, Lepage:2020tgj}.

We have first brute-forcibly computed $\F{000}(x)$ over the full three-dimensional
domain. This approach proves to be quite inefficient with relatively
large error bars, but its real part has been reported and referred to
as VEGAS 3D in Fig.~\ref{fig:divergent}.

Another more accurate method is to remark that the integral $\F{0}(x)$
of Eq.~\eqref{eq:F0} can be analytically performed, thus reducing the
integration dimensionality explored by VEGAS to two dimensions
(referred to as VEGAS 2D in Fig.~\ref{fig:divergent}). Indeed, one
has~\cite{2007tisp.book.....G}
\begin{equation}
  \F{0}(x) = \Eone(-2ix) = -\Ei(2i x) + i \pi
\end{equation}
where $\Eone$ and $\Ei$ are the Exponential integral functions. Note
that this expression disagrees with Eq.~(B.1) in
\citet{Ballardini:2024irx}. The reason is that $\Ei$ is defined on
the real domain, but the continuation to the complex plane can be
ambiguous due to singularities at $0$ and $\infty$. We have
consistently used the \texttt{scipy} implementation of the exponential
integral and one can easily check that $\Ei(\pm 2i x) \to \pm i\pi$.
A factor $i\pi$ is therefore necessary to recover the expected
behavior of $\F{0}(x)$ at infinity. This problem does not occur if one
uses $\Eone$ instead of $\Ei$.

Since the disagreement concerns only the finite part of $\F{000}(x)$,
we can subtract from the numerical integration the known diverging
part when $x \to 0$. From Eq.~\eqref{eq:exactF000}, we define
\begin{equation}
  \F{000}^\mathrm{div}(x) \equiv - \left(\frac{\pi^2}{4} +
  \frac{B^2}{2}\right) \ln(x) - \frac{B}{2} \ln^2(x) - \frac{1}{6}
  \ln^3(x).
\end{equation}
Fig.~\ref{fig:divergent} shows the real and imaginary parts of
$\F{000}(x) - \F{000}^\mathrm{div}(x)$ obtained by the numerical
integrations in two and three-dimensions and we find
\begin{equation}
\calZ_\mathrm{vegas}= (2.8051 \pm 0.0014)  + (0.00089\pm 0.00096)i.
\end{equation}
Both the real and imaginary parts match, and only match, the value of
$\calZ = 7 \zeta(3)/3$ that we found in \citet{Auclair:2022yxs}. The
two values of $\calZ$ proposed in \citet{Ballardini:2024irx} are
excluded: $\calZ = \zeta(3)/3$ fails to be consistent with the real
part, and $\calZ\simeq2.97353-0.0273557 i$ gives an incorrect real
\emph{and} imaginary part.

\section{Conclusion}

In conclusion, the right form for $\F{000}(x)$ is given by
Eq.~\eqref{eq:exactF000} and this implies that the correct expression
for the $\bsdiam{i}$ and $\btdiam{i}$ appearing in the power
spectra of Eqs.~\eqref{eq:calPz} and \eqref{eq:calPh} should be taken
from \citet{Auclair:2022yxs}.

\section*{Acknowledgement}

We would like to warmly thank J\'er\^ome Martin and Patrick Peter for
professional advices and encouragements. CR thanks the Institut
d'Astrophysique de Paris for hosting and support. This work is also
supported by the ESA Belgian Federal PRODEX Grants $\mathrm{N^{\circ}}
4000143201$ and $\mathrm{N^{\circ}} 4000144768$.

\bibliographystyle{elsarticle-num-names}

\bibliography{biblio}

\end{document}